\documentclass[%
aps,
preprint,
 amsmath, amssymb]{revtex4-1}

\usepackage[english]{babel}
\usepackage{graphicx}% Include figure files
\usepackage{dcolumn}% Align table columns on decimal point
\usepackage{bm}% bold math
\usepackage{textcomp}
\usepackage{color}
\usepackage{natbib} % Use modern font encodings

\begin{document}
\title{Bandgap engineering in an epitaxial two-dimensional  honeycomb Si$_{6-x}$Ge$_x$  alloy}

\author{A. Fleurence} \email{antoine@jaist.ac.jp}\affiliation{School of Materials Science, Japan Advanced Institute of Science and Technology, 1-1 Asahidai Ishikawa  923-1292, Japan}
\author{Y. Awatani}\affiliation{School of Materials Science, Japan Advanced Institute of Science and Technology, 1-1 Asahidai Ishikawa  923-1292, Japan}
\author{C. Huet}\affiliation{School of Materials Science, Japan Advanced Institute of Science and Technology, 1-1 Asahidai Ishikawa  923-1292, Japan}
\author{F. B. Wiggers} \affiliation{MESA+ Institute for Nanotechnology, University of Twente, 7500 AE Enschede, The Netherlands}\altaffiliation{Current address: ASML Netherlands B. V.,  De Run 6501, 5504 DR Verdhoven,  The Netherlands}
\author{S. M. Wallace}\affiliation{School of Materials Science, Japan Advanced Institute of Science and Technology, 1-1 Asahidai Ishikawa  923-1292, Japan} \altaffiliation{Current address: International Center for Materials Nanoarchitectonics, National Institute for Materials Science, Tsukuba, 305-0044, Japan and the Graduate School of Pure and Applied Sciences, University of Tsukuba, Tsukuba, 305-8573, Japan.}
\author{T. Yonezawa}\affiliation{School of Materials Science, Japan Advanced Institute of Science and Technology, 1-1 Asahidai Ishikawa  923-1292, Japan}
\author{Y. Yamada-Takamura}\affiliation{School of Materials Science, Japan Advanced Institute of Science and Technology, 1-1 Asahidai Ishikawa  923-1292, Japan}

%\abbreviations{IR,NMR,UV}
%\keywords{Silicene, SiGe Alloy, Bandgap enginnering, 2D materials, STM, ARPES, DFT}

\begin{abstract}
  In this Letter, we demonstrate that it is possible to form a  two-dimensional (2D) silicene-like Si$_5$Ge compound by replacing the Si atoms occupying  on-top sites in the planar-like structure of epitaxial silicene   on ZrB$_2$(0001)  by deposited Ge atoms. For coverages below 1/6 ML, the Ge deposition gives rise to  a Si$_{6-x}$Ge$_{x}$ alloy (with $x$ between 0 and 1) in which the on-top sites are randomly occupied by Si or Ge atoms.  The progressive increase  of the valence band maximum with $x$ observed experimentally  originates from a selective charge transfer from Ge atoms to  Si atoms. These achievements provide evidence for the possibility of engineering the bandgap in  2D SiGe alloys in a way that is similar for their bulk counterpart.
\end{abstract}

\maketitle
% \section{Introduction}      
 Alloying materials with similar structures and miscible elements is of great interest for a wide range of applications as it  allows for  adjusting various  parameters  to values which can not be  achieved with  elemental materials or compounds.  This versatility is well exemplified by the engineering of the bandgap  of semiconducting  alloys which makes possible the  fine tuning of the wavelength of solid-state lightings   by controlling the alloys composition. With the continuous efforts to scale down the dimension of elementary bricks of electronic devices, the fabrication of low-dimensional  alloys, including two-dimensional (2D) materials, became  a technologically important challenge \cite{Ning17} as it was for bulk semiconducting materials in the past.  Alloying  semimetallic graphene, with its isomorphic  wide-bandgap  analogue h-BN  which would have permitted to set the value of  the bandgap of a 2D h-BNC alloy in a wide energy range,  was however found to be hindered by the low miscibility of the two materials resulting in a phase segregation \cite{Ci10}. In contrast, ternary and quaternary alloys of transition metal dichalcogenide   could be synthethized successfully \cite{Komsa12,Su14, Fu15,Duerloo15,Susarla17} and the  tunability of the optical bandgap was demonstrated. 
Among the 2D materials experimentally fabricated, silicene, a 2D honeycomb latttice of Si atoms,  has the particularity to  allow for  continuing   scaling down the Si-based nanoelectronics \cite {Tao15}.  Thorough efforts were put into evaluating methods for tuning the electronic properties  of silicene including doping \cite{Sahin13, Friedlein13}, or the adsorbtion of adatoms or molecules \cite{Houssa11, Osborn11, Huang13}. Alloying free-standing forms  of silicene and   germanene, its Ge analogue,  investigated  by first principles calculations \cite{Sahin09,Padilha13,Zhang14} suggested that such 2D hexagonal SiGe alloys are  stable and  various parameters including the lattice parameter or the spin-orbit gap open in the Dirac cones were found to be tunable with  the  Si:Ge ratio while the non-triviality of the band structure is preserved.\\
 In this Letter, we report the realization of a 2D SiGe epitaxial alloy fabricated by depositing Ge on silicene on zirconium diboride (ZrB$_2$) films grown on Si(111) \cite{Fleurence14}. Furthermore, we investigated  the possibility of engineering its bandgap  in a way similar to  bulk SiGe alloys.\\
%\section{Experimental and Computational}
Epitaxial silicene sheets were prepared by annealing  ZrB$_2$ thin films epitaxially grown on Si(111)  \cite{Fleurence12,YYT10} in ultrahigh vacuum (UHV). The deposition of Ge on silicene was realized by means of a Knudsen cell implemented in each of the UHV systems used for these experiments. The Ge flux, calibrated in each of these systems by depositing Ge on a Si(111) substrate, was in the 0.09 - 0.12 ML.min$^{-1}$ range (1 monolayer (ML) refer to the density of atoms in epitaxial silicene on ZrB$_2$(0001): $1.73 \times10^{15} at.cm^{-2}$). Scanning tunneling microscopy (STM)  was performed at room temperature. Photoemission spectroscopy experiments were conducted at beamline BU06 of UVSOR. Core-level spectra in normal emission  and  angle-resolved photoemission spectroscopy (ARPES) spectra were recorded at room temperature and at  20 K, respectively. The respective energy resolutions as estimated from the broadening of the Fermi edge are 35 and 10 meV.\\
DFT calculations within a generalized gradient approximation (GGA) \cite{Kohn65,Perdew96} were performed using the OPENMX code \cite{Ozaki03} based on norm-conserving pseudopotentials generated with multireference energies \cite{Perdew96} and optimized pseudoatomic basis functions \cite{Ozaki03}. The two input structures consist of ($2\times2$) ZrB$_2$(0001) slabs  made of 8 Zr and 7 B layers terminated on both face respectively by silicene  or Si$_5$Ge layers. A 42 \AA \space vacuum space is separating the slabs.   For Zr atoms, a s3p2d2 basis function i.e. including  three, two, and two optimized radial functions allocated respectively to  the s, p, and d orbitals. For  Si, Ge  and B atoms, s2p2d1 basis functions were adopted. A cutoff radius of 7 Bohr was chosen for all the basis functions. A regular mesh of 220 Ry in real space was used for the numerical integrations and for the solution of the Poisson equation. A (5$\times$5$\times$1) mesh of k points was used. For geometrical optimization, the force on each atom was relaxed to be less than 0.0001 Hartree/Bohr. In order to take into account  the strength of translational symmetry breaking, the spectral weight as derived from the imaginary part of the one-particle Kohn-Sham Green function, was unfolded to the  Brillouin zone of the "one-Si-atom unit cell"   \cite{Lee14}  following a method described in Ref. \cite{Lee13_unfolding}.\\
%\section{Results}

\begin{figure}[h!]
	\centering
\includegraphics[width=8cm]{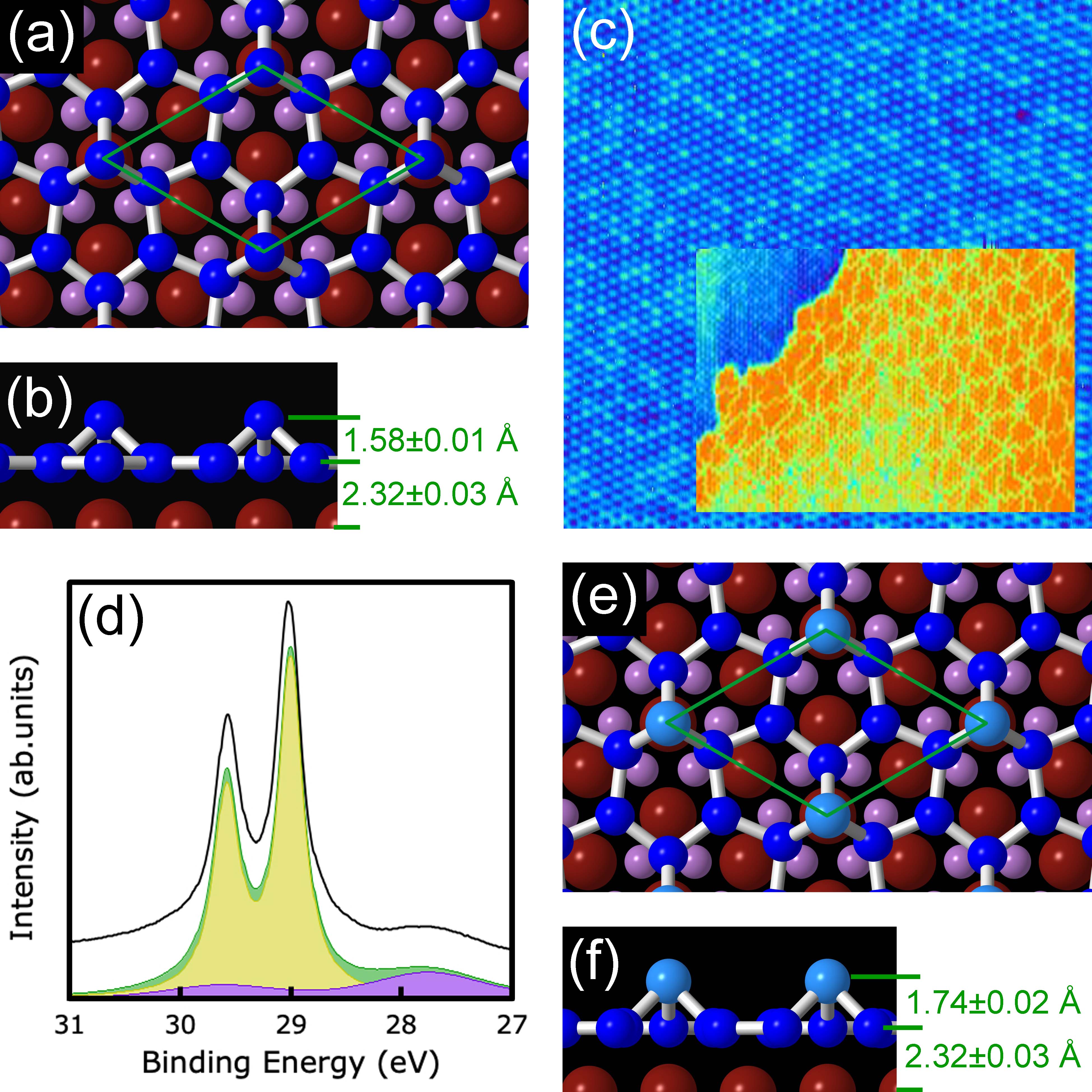}
\caption{\textbf{Deposition of Ge on silicene on ZrB$_2$(0001).} (a) and (b): Top- and side-views of the  epitaxial planar-like structure of silicene on ZrB$_2$(0001) as determined by DFT calculations. Si, Zr and B atoms are respectively dark blue-, red- and purple-colored. The green rhombus indicate the ($\sqrt{3}\times\sqrt{3}$)-reconstructed unitcell.  (c): STM image (30 nm$\times$ 30 nm, $V$=1.0 V, $I$=100 pA) of epitaxial silicene after deposition of 0.057 ML Ge. The STM image (33 nm$\times$23 nm) in the  inset  shows  the  silicene-Ge alloy (top left) and Si bilayer islands side-by-side. (d): Photoemission spectrum recorded with h$\nu$=80 eV in the Ge3d and Zr4p region after deposition of 0.090 ML Ge.  The experimental spectrum  is  indicated by the black line.  Yellow- and purple-filled curves are the contribution of Ge3d and Zr 4p core-levels determined by fitting  and the green-filled curve is their sum.  The full widths at half maximum are 270 meV and 290 meV respectively for the Ge3d$_{3/2}$ and Ge3d$_{5/2}$ peaks. (e) and (f): Top- and side-views of the structure of epitaxial Si$_5$Ge on ZrB$_2$(0001) as determined by DFT calculations. Si, Zr and B  atoms are  colored in the same way as in Figure (a)  and Ge atoms are light blue-colored. }
\label{Fig1}
\end{figure}

\begin{figure}[h!]
	\centering
	\includegraphics[width=8 cm]{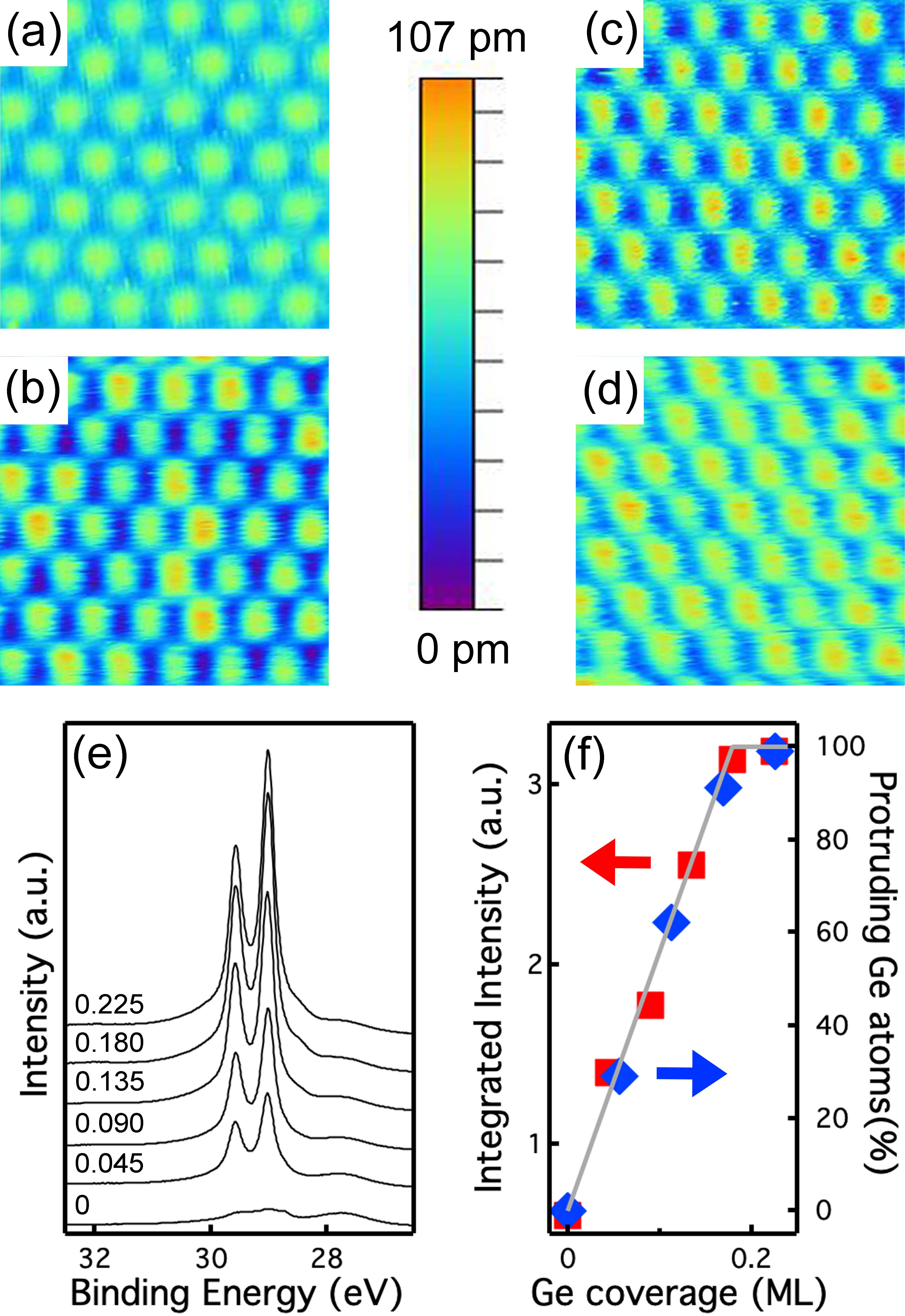}
		\caption{ \textbf{Si$_{6-x}$Ge$_x$ alloy} (a)-(d): STM images (4 nm$\times$ 4 nm, $V$=1.0 V,  $I$=100 pA) after deposition of 0.030 ML Si, 0.057, 0.113, and 0.167 ML Ge. Their common color-coded z scalebar is shown. (e) Spectra recorded in the Ge 3d and Zr4p core-levels region recorded for different Ge coverage with a photon energy of 80 eV. (f):  Integrated intensity of the photoemission spectra  and   percentage of protruding Ge atoms as functions of the Ge coverage.}
		\label{Fig2}
\end{figure}

\begin{figure}[h!]
	\centering
	\includegraphics[width=17 cm]{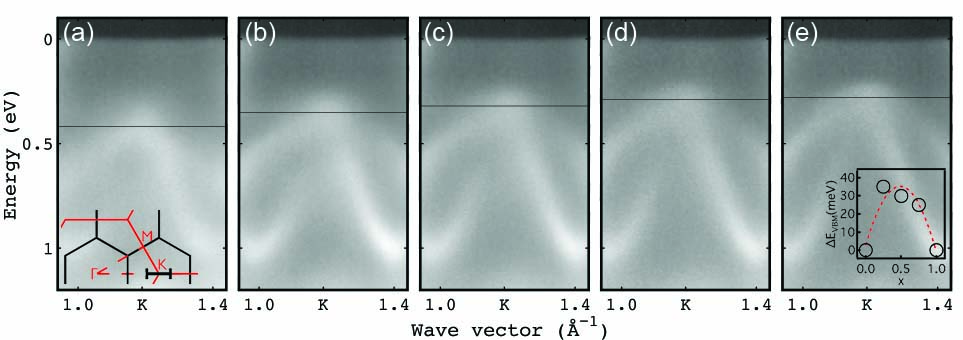}
		\caption{\textbf{Valence band of Si$_{6-x}$Ge$_x$ alloy.} The inset of the figure  (a) shows the region around the $K$ point of the Brillouin zone of unreconstructed silicene  in which the spectra were recorded.    Brillouin zones of  ($\sqrt{3} \times \sqrt{3}$)-reconstructed and unreconstructed silicene are respectively indicated  by  black and red lines.  (a)-(e): ARPES spectra  recorded with a photon energy of h$\nu$=45 eV on Si$_{6-x}$Ge$_x$ alloys   for $x$ = 0, 0.25, 0.5, 0.75 and 1. The horizontal lines indicate $E_{VBM}$.  The inset of Fig. (e) shows $\Delta E_{VBM}=E_{VBM}-(E_{VBM}^{Silicene}x+  E_{VBM}^{Si_{5}Ge}(1-x))$ as a function of $x$. Its fitting with $b x(1-x)$ and $b$ =-135 meV is indicated by a dashed red line. } 
\label{Fig3}
\end{figure}

\begin{figure}[h!]
	\centering
	\includegraphics[width=17 cm]{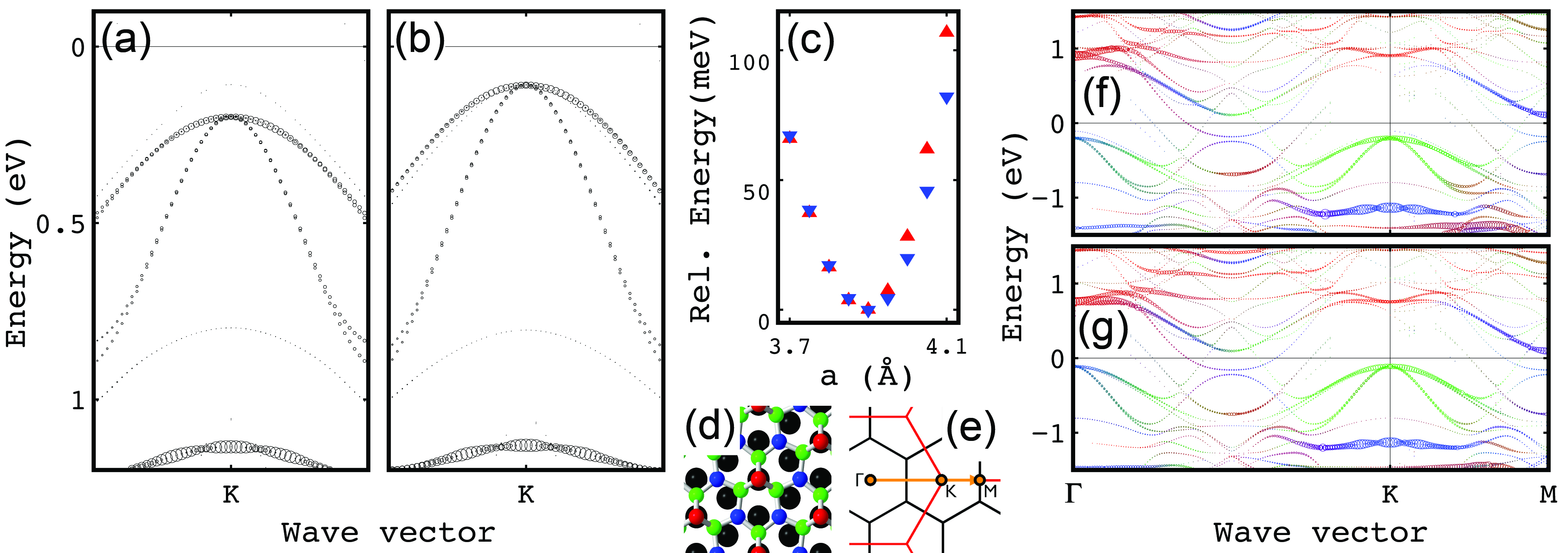}
		\caption{\textbf{Band structure of silicene and Si$_5$Ge}  (a) and (b): Band structures of  silicene and Si$_5$Ge in the region indicated in the inset of Fig. \ref{Fig3} (a).  The spectral weight of the combined contribution of Si and Ge atoms is indicated by the size of the circles.  (c): Lattice parameter-dependence of the energy per atom of free-standing planar-like structures of  silicene (red) and  Si$_5$Ge (blue). (d): Schematics of the planar-like structure on Zr-terminated ZrB$_2$(0001).  Zr,  on-top, bridge and hollow atoms are respectively black-, red-, green- and blue-colored. (e). Schematics of the k-space. Black and red lines indicate the ($\sqrt{3}\times\sqrt{3}$)  and ($1\times1$)  Brillouin zones of silicene.   (f) and (g): Calculated band structures for silicene and Si$_5$Ge along the path indicated in Fig. (e). The contribution of the  on-top, bridge and hollow sites atoms are respectively red-, green- and blue-colored in agreement with Fig. (d). The size of the circles represents the spectral weight of the combined contribution of Si and Ge atoms. }
\label{Fig4}
\end{figure}

Silicene crystallises spontaneously on ZrB$_2$(0001) in a so called ``planar-like" ($\sqrt{3}\times\sqrt{3}$)-reconstructed  structure \cite{Lee13,Lee14} adopted by several forms of epitaxial silicene  \cite{Chen12,Ming13,Aizawa14}.  This structure  fits with the ($2\times 2$) unitcell of ZrB$_2$(0001) in such a way that $a_{Si} = \frac{2}{\sqrt3}a_{ZrB_2}$, where $a_{ZrB_2}$ (3.178 \AA) and $a_{Si}$  (3.65 \AA) are   the lattice parameters of unreconstructed silicene and  ZrB$_2$(0001), respectively. The Figs. \ref{Fig1}.(a)  and (b) show the details of the planar-like structure  as the result of optimization in DFT calculations.  In this structure, two, three and one Si atoms are respectively sitting on hollow, bridge and on-top sites of the Zr-terminated thin films.  All of the Si atoms but one are laying 2.3 \AA \space above the  topmost Zr layer whereas the  Si atoms  sitting on the on-top sites visible in STM images are protruding  at   3.9 \AA. As shown in the STM image of Fig. \ref{Fig1}.(c), the deposition of 0.05 ML Ge on silicene turns the domain structure of the pristine silicene sheet \cite{Fleurence12, Nogami19} into a single domain in a way similar to the deposition of silicon\cite{Fleurence16}. However, in contrast to  silicon atoms,  the deposition of Ge atoms results in a contrast between the protrusions,  observed  for all scanning conditions,  and  most visible  for a bias voltage of 1.0 V, which suggests that some Ge atoms substituted Si protruding atoms.   As this Ge coverage is beyond that required to fully turn the domain structure of silicene into a single-domain (0.03 ML) \cite{Fleurence16},   the excess of atoms results  locally in the formation of bilayer silicon  islands \cite{Gill17} like the one shown in the inset of Fig. \ref{Fig1}.(c). These islands are rare  and distant (few hundreds  of nanometers from each others).\\ 
 The Fig. \ref{Fig1}.(d) shows a photoemission spectrum recorded  with a photon energy of  h$\nu$=80 eV in the region of the Zr4p and Ge3d core-levels  after deposition of 0.09 ML Ge. The fact that the Ge3d component  can be fitted with a single pair of Lorentzian functions, points out that the Ge atoms  are incorporated into a single site, i.e.  the on-top sites of the silicene lattice. \\
The Fig. \ref{Fig1}.(d) presents such a planar-like structure   after  optimization  by DFT, which appears to be essentially similar to that  shown in Fig. \ref{Fig1}.(a). The main difference is  the length of the bonds between atoms of the on-top and bridges sites  which increases  from 2.37 \AA \space  to 2.47 \AA. This distance is  longer than that of the Si-Ge bonds measured in bulk SiGe alloys \cite{Aubry99}  or calculated for 2D hexagonal SiGe alloys \cite{Padilha13}. The Ge atom is located  1.74 \AA \space above the bottom Si atoms instead of 1.58 \AA \space for the on-top Si atom in silicene which confirms that the  taller protrusions  must  be assigned to Ge atoms. 
Fig. \ref{Fig2} shows  the evolution with the Ge coverage of the silicene sheet as imaged by STM  and photoemission spectra recorded in the Zr4p and Ge3d  core-levels region. The number of protruding Ge atoms and the integrated intensity of the spectrum  increase both  linearly until a Ge coverage of  0.17 ML close to the density of protruding atoms in the planar-like  structure (1/6 ML)  is reached.  One can deduce that  below this coverage, Ge adatoms are fully incorporated into the silicene sheet and replace systematically protruding Si atoms in the planar-like structure. This shows that it is possible to fabricate a 2D  Si$_{6-x}$Ge$_x$  alloy with  $x$ being finely adjustable  between 0  and  1  by depositing a controlled amount of Ge in this  coverage range. \\
To determine the effect of the Ge atoms  on the band structure of Si$_{6-x}$Ge$_x$, ARPES spectra were recorded   for different values of $x$  between 0 and 1. The Fig. \ref{Fig3}   show  spectra recorded  with a photon energy of 45 eV in the region of the  $K$ point of the Brillouin zone of unreconstructed silicene where the  valence band maximum (VBM)  is located   \cite{Fleurence12, Lee13,Lee14}.   One can see that the top of the binding energy of the valence band  E$_{VBM}$ evolves steadily from  $E_{VBM}^{Silicene}$=0.42 eV for silicene  to  $E_{VBM}^{Si_{5}Ge}$=0.28 eV for Si$_5$Ge whereas the bottom of the band remains at a binding energy of 1.0 eV.  The evolution of E$_{VBM}$ with  $x$ is not linear and the fitting of   $\Delta E_{VBM}=E_{VBM}(x)-(E_{VBM}^{Silicene}x+ E_{VBM}^{Si_{5}Ge}(1-x))$ with  $b x(1-x)$ gives a bowing parameter $b$ of -135 meV (See inset of Fig. \ref{Fig3}.(e)). \\
 As shown in Figs. 4.(a) and (b), the difference in  band structure between silicene and $Si_{5}Ge$ observed experimentally  is well reproduced by DFT calculations. The band structures were calculated for structures with a lattice parameter artificially increased by 5 \% \cite{Lee14}  to compensate the overestimation of the bandwidth caused by the GGA \cite{Eastman80,Haas09}.  In agreement with the experimental ARPES spectra,    the VBM is shifted upwards by 90 meV (from 200 meV to 110 meV) whereas the bottom of the band remains at the same energy. \\ 
To evaluate the influence of the epitaxial strain on the band structure of the  Si$_{6-x}$Ge$_{x}$ alloy,  the energy of  the free-standing planar-like structures of silicene and Si$_5$Ge  were calculated as a function of the lattice parameter of the unreconstructed silicene structure.   To preserve the  planar-like structure upon geometrical optimization in absence of the subtrate, the  5 Si atoms of the bottom layers  were forced to remain in the same plane.   In contrast  to the slight increase of the equilibrium lattice parameter found for free-standing  2D hexagonal SiGe alloy \cite{Padilha13},  the planar-like structures of silicene and Si$_5$Ge have the same equilibrium  lattice parameter of 3.89 \AA \space (Fig. \ref{Fig4}.(c)), corresponding to a compressive strain of 6.2 \% which suggests that any strain-induced effect on evolution of  the band structure is negligible. \\
 The good agreement between  experimental and computed band structures allows for analysing further the nature of the effect of the Ge atoms. The Figs. \ref{Fig4}.(f) and (g) show the respective contributions of on-top, hollow and bridges sites atoms,  as indicated in Fig. \ref{Fig4}.(d), to  the band structures of  epitaxial silicene and Si$_5$Ge plotted along the path indicated in Fig. \ref{Fig4}.(e). For both structures, the valence band centered on the $K$ point appears to originate from the Si atoms in the bridge sites, whereas the conduction band minimum (CBM) centered on the $M$ point of the Brillouin zone of the unreconstructed silicene originates from the Si atoms of  the hollow sites. One can observe that in contrast to  $E_{VBM}$, $E_{CBM}$ the energy of the CBM does not vary much between the two structures.  
  The comparison of the computed Mulliken charges carried by the different  atoms  (Table \ref{Mulliken}) suggests that the on-top site Ge atoms are electron richer than the Si atoms in the same position, in agreement with the higher  electronegativity of Ge (2.01) in comparison to that of Si (1.90). This induces an increase of  the  electron donation from the bridge site atoms which are the first neighbors of the on-top site atoms and thus become further positively charged. In contrast, the charge carried by the hollow-sites atoms does not vary significantly.  This selective charge transfer from on-top sites atoms to bridge sites atoms results in a progressive shift of $E_{VBM}$ towards the Fermi level whereas $E_{CBM}$ is fixed. \\ 
  
  In conclusion, we experimentally demonstrated the  possibility of fabricating an epitaxial silicene-like Si$_5$Ge compound by depositing a minute amount of Ge on silicene on ZrB$_2$(0001) thin films on Si(111). For Ge coverages below 1/6 ML, the deposition gives rise to a  Si$_{6-x}$Ge$_x$ alloy  based on the planar-like structure of epitaxial silicene in which the protruding sites are randomly occupied by Si or Ge  atoms. The substitution of Si atoms by Ge atoms  induces a shift of the VBM which allows for finely tuning the bandgap of the epitaxial Si$_{6-x}$Ge$_x$ by controlling the amount of Ge atoms.\\ %The  selective charge transfer   from the Ge atoms and Si atoms which makes possible a bandgap engineering in Si$_{6-x}$Ge$_x$.% including  transition metal dichalcogenide single layers.  

%\section{acknowledgement}

A.F. acknowledges financial support from JSPS KAKENHI Grant Number 26790005. F.B.W. acknowledges financial support from the Foundation for Fundamental Research on Matter (FOM, Project No. 12PR3054), which is part of the Netherlands Organization for Scientific Research (NWO).
 A part of this work was supported by the Joint Studies Program of the  Institute for Molecular Science No. 203 (2016-2017) and No. 206 (2017-2018).  This work was financially supported by  
Iketani Science and Technology Foundation. We are thankful to Dr. Hiroyuki Yamane  (IMS) for his assistance at beamline BL6U of UVSOR.

\newpage 

%\bibliographystyle{apsrev4-1}
%\bibliography{Silicene_Ge}

%

\newpage

\newpage

\newpage

\newpage

\newpage

 \begin{table}[h]
\begin{tabular}{|c|c|c|c|}
\hline
 & On-top & Bridge   & Hollow  \\
\hline \hline
Silicene  & -0.052 &0.083 &  -0.016   \\
\hline
Si$_5$Ge  & -0.711 &  0.322    & -0.020   \\
\hline
\end{tabular}
\caption{\textbf{Mulliken charges expressed in number of e$^-$ calculated by DFT}.}
\label{Mulliken}
\end{table}

\end{document}